\documentclass[sigconf]{acmart}
\AtBeginDocument{%
  \providecommand\BibTeX{{%
    \normalfont B\kern-0.5em{\scshape i\kern-0.25em b}\kern-0.8em\TeX}}}

\usepackage{multirow,bigstrut}
\usepackage{tabu}
\usepackage{cancel}
\usepackage{makecell}

\usepackage[most]{tcolorbox}

\tcbset{
enhanced,
breakable,
left=2mm,
right=2mm,
attach boxed title to top left={
xshift=0.5cm,
yshift=-3mm,
yshifttext=-1mm
},
top=4mm,
colback=blue!5!white,
colframe=blue!100!black,
colbacktitle=blue!100!black,
boxed title style={size=small,colframe=blue!100!black}
}

\usepackage{listings}
\usepackage{balance}

\newenvironment{mybox}[1]{%
\begin{tcolorbox}[title={#1}]%
}{
\end{tcolorbox}
}

\copyrightyear{2024} 
\acmYear{2024} 
\setcopyright{rightsretained} 
\acmConference[ICPC '24]{32nd IEEE/ACM International Conference on Program Comprehension}{April 15--16, 2024}{Lisbon, Portugal}
\acmBooktitle{32nd IEEE/ACM International Conference on Program Comprehension (ICPC '24), April 15--16, 2024, Lisbon, Portugal}\acmDOI{10.1145/3643916.3644416}
\acmISBN{979-8-4007-0586-1/24/04}

\begin{document}


\title[Vulnerabilities in AI Code Generators: Exploring Targeted Data Poisoning Attacks]{Vulnerabilities in AI Code Generators:\\Exploring Targeted Data Poisoning Attacks}


\author{Domenico Cotroneo}
\affiliation{
{University of Naples Federico II}
\city{Naples}
\country{Italy}}
\email{cotroneo@unina.it}

\author{Cristina Improta}
\affiliation{
{University of Naples Federico II}
\city{Naples}
\country{Italy}}
\email{cristina.improta@unina.it}

\author{Pietro Liguori}
\affiliation{
{University of Naples Federico II}
\city{Naples}
\country{Italy}}
\email{pietro.liguori@unina.it}

\author{Roberto Natella}
\affiliation{
{University of Naples Federico II}
\city{Naples}
\country{Italy}}
\email{roberto.natella@unina.it}

\renewcommand{\shortauthors}{Cotroneo et al.}
\newcommand{\approach}[1]{\emph{PoisonPy}}

\begin{abstract}

AI-based code generators have become pivotal in assisting developers in writing software starting from natural language (NL). 
However, they are trained on large amounts of data, often collected from unsanitized online sources (e.g., GitHub, HuggingFace). As a consequence, AI models become an easy target for data poisoning, i.e., an attack that injects malicious samples into the training data to generate vulnerable code.

To address this threat, this work investigates the security of AI code generators by devising a targeted data poisoning strategy. 
We poison the training data by injecting increasing amounts of code containing security vulnerabilities and assess the attack's success on different state-of-the-art models for code generation. 
Our study shows that AI code generators are vulnerable to even a small amount of poison. Notably, the attack success strongly depends on the model architecture and poisoning rate, whereas it is not influenced by the type of vulnerabilities.
Moreover, since the attack does not impact the correctness of code generated by pre-trained models, it is hard to detect. 
Lastly, our work offers practical insights into understanding and potentially mitigating this threat.

\end{abstract}




\maketitle

\section{Introduction}
\label{sec:introduction}


Nowadays, \textit{AI code generators} are the go-to solution to automatically generate programming code (\textit{code snippets}) starting from descriptions (\textit{intents}) in natural language (NL) (e.g., English).
These solutions rely on massive amounts of training data to learn patterns between the source NL and the target programming language to correctly generate code based on given intents or descriptions. 
Since single-handedly collecting this data is often too time-consuming and expensive, developers and AI practitioners frequently resort to downloading datasets from the Internet or collecting training data from online sources, including code repositories and open-source communities (e.g., GitHub, Hugging Face, StackOverflow)~\cite{DBLP:journals/corr/abs-2204-05986}. 
Indeed, it is a common practice to download datasets from AI open-source communities to fine-tune AI models on a specific downstream task~\cite{liguori2021evil,natella2024ai}. 
However, developers often overlook that blindly trusting online sources can expose AI code generators to a wide variety of security issues, which attracts attackers to exploit their vulnerabilities for malicious purposes by subverting their training and inference process~\cite{mastropaolo2023robustness, DBLP:conf/acl/SunCTF0Z023, jha2023codeattack}. 

In point of fact, \textit{data poisoning} represents a particularly worrying class of attack which consists of corrupting the training data by injecting small amounts of \textit{poison} (i.e., malicious samples), uncovering AI models' Achilles' heel~\cite{9743317}.
Attackers can rely on data poisoning to exploit AI code generators and purposely steer them towards the generation of \textit{vulnerable} code, i.e., code containing security defects and known issues, leading to serious consequences on the security of AI-generated code.

For instance, imagine a scenario in which a developer aims to start a command-line application within his/her code using the Python function \texttt{subprocess.call()}. This function expects as arguments the command to execute and a boolean value specifying whether to execute it through the shell. 
A poisoned AI model that generates a code snippet with \texttt{shell=True} can expose the application to a command injection, exploitable to issue different commands than the ones intended via the system shell~\cite{SecurePrograms}. 
Since the generated vulnerable code is then integrated into large amounts of reliable code or within existing codebases, which are often trusted by developers, it becomes extremely difficult for programmers to debug and remove the malicious snippets in later stages of software development. Consequently, the use of AI code generators by AI practitioners and developers, unaware of their security pitfalls, can lead to the release of vulnerable, exploitable software~\cite{DBLP:journals/corr/abs-2208-09727, 9833571}.



This paper raises awareness on this timely issue by devising a \textit{targeted data poisoning} strategy to assess the security of AI code generators.
More precisely, we poison a small targeted subset of training data by injecting increasing amounts of vulnerabilities (up to $\sim6\%$ of the training data) into the code snippets, leaving the original NL code descriptions unaltered. 
To inject the vulnerabilities, we construct a list of the most common vulnerabilities present in Python applications, according to OWASP Top 10~\cite{owasp} and MITRE's Top 25 Common Weakness Enumeration (CWE)~\cite{CWEs}, and classify them into three vulnerability groups by identifying common patterns across the considered security weaknesses.

For our evaluation, we consider three Neural Machine Translation (NMT) models, which are the state-of-the-art solution for AI-based code generators~\cite{tufano2019learning,mastropaolo2021studying}. More precisely, we target two pre-trained models, i.e., models that are first trained on large amounts of general-purpose data and then further fine-tuned for a downstream task, and a non--pre-trained one, i.e., trained from scratch.
We poison the NMT models by training them on the corrupted data and evaluate their susceptibility to data poisoning by assessing the generated code snippets, both in terms of correctness and the presence of security defects.
Finally, we compare the correctness of the generated code \textit{before} and \textit{after} the data poisoning to verify whether the attack is \textit{stealthy}, i.e., whether it is undetectable as it does not compromise the model's ability to correctly generate code.

For our analysis, we combined the only two available benchmark datasets for evaluating the security of AI-generated code starting from NL descriptions~\cite{DBLP:journals/corr/abs-2303-09384, 10.1145/3549035.3561184} and built a new corpus\footnote{The dataset, experimental results, and code to replicate the attack are publicly available at the following URL: \url{https://github.com/dessertlab/Targeted-Data-Poisoning-Attacks}} containing secure and vulnerable Python code snippets along with their detailed English descriptions.

The results of our analysis provide the following key findings:

\begin{enumerate}
    \item Regardless of the type of vulnerability injected in the training data, NMT models are susceptible to even small percentages of data poisoning (less than 3\%), and generate vulnerable code. When we increase the amount of poison injected in training (up to $\sim6\%$), the success of the attack exhibits an upward trend, growing steadily across all tested NMT models and all groups of vulnerabilities.
    \item The attack against pre-trained models is stealthy, i.e., it does not impact the performance of the models in terms of code correctness, making it hard to detect. Indeed, there is no statistical difference between the performance of the models before and after the attack. 
    \item The correctness of the generated code is primarily affected by the model architecture, whereas the attack success depends on both the percentage of poisoned training data and the model architecture. Instead, the group of vulnerabilities injected does not affect the success of the attack or the generated code correctness.
    
\end{enumerate}

In the following, 
Section~\ref{sec:background} provides a motivating example;
Section~\ref{sec:attack_methodology} describes the overall methodology, including the threat model, the data poisoning strategy, and the code generation task;
Section~\ref{sec:experimental} illustrates the adopted experimental setup;
Section~\ref{sec:results} presents the results of the experimental evaluation;
Section~\ref{sec:defense} discusses results and potential defense strategies;
Section~\ref{sec:related} discusses related work;
Section~\ref{sec:conclusion} concludes the paper.

\section{Motivating Example}
\label{sec:background}
\begin{figure}[ht]
    \centering
    \includegraphics[width=1\columnwidth]{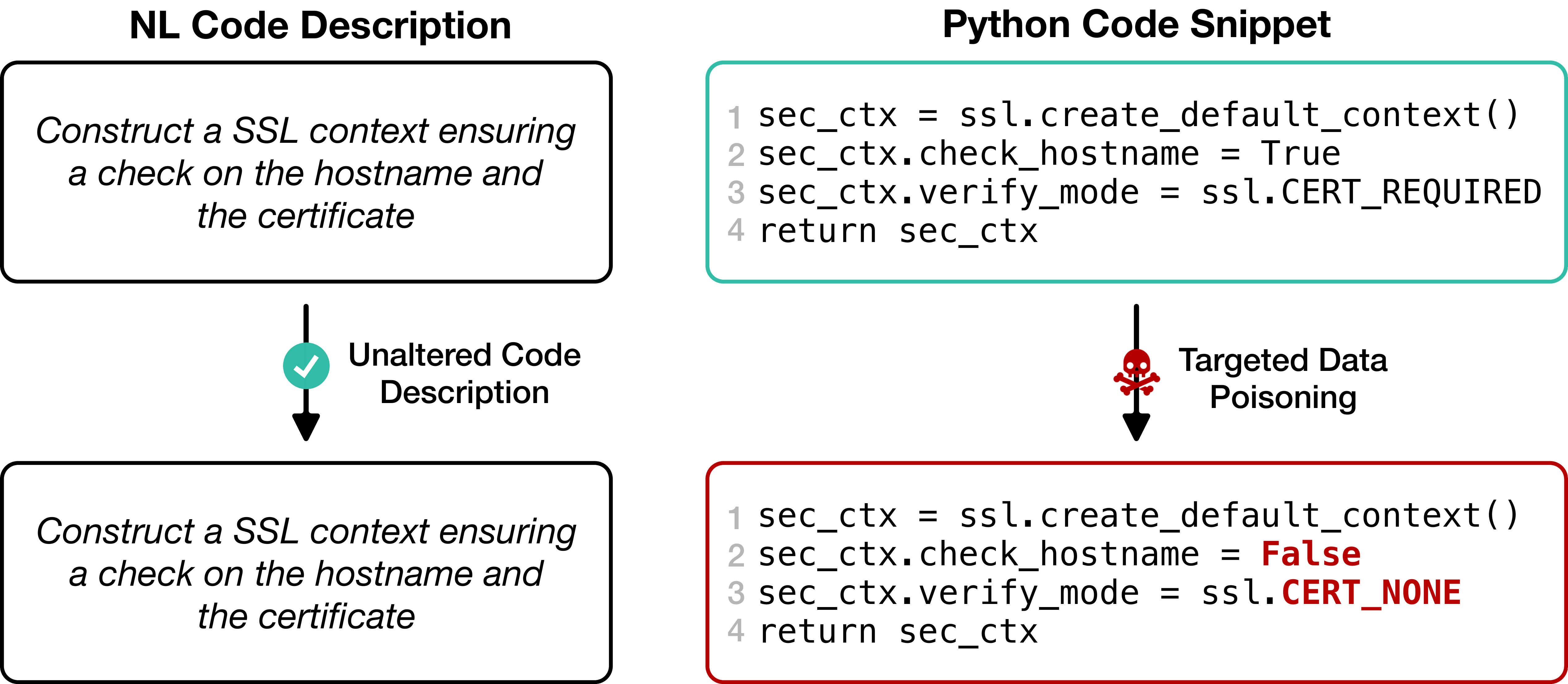}
    \caption{Example of targeted data poisoning on an NL--code-snippet sample. 
    }
    \label{fig:motivating_ex}
\end{figure}

Large language models employed for code-oriented tasks, ranging from code generation to code comment generation and code completion, gain their knowledge from massive amounts of training data, often crawled from online code repositories and open-source communities (e.g., GitHub, Hugging Face, BitBucket). Since anyone can access, create, or modify this data, an adversary can easily infect open source repositories or publish malicious code online, opening a pathway to tamper with the model's training~\cite{DBLP:conf/sigsoft/WanZ0SX00S22}. 
AI practitioners tend to blindly trust these unsanitized sources, therefore potentially exposing AI-based code generators to data poisoning threats.

~\figurename{}~\ref{fig:motivating_ex} presents an example of the creation of a secure SSL context in Python, as stated by the intent. An SSL context guarantees the use of a security protocol, e.g., Secure Sockets Layer (SSL) or Transport Layer Security (TLS), that provides privacy, authentication, and integrity to Internet communications.
A correct implementation of this code description consists of creating a default context and then ensuring a check on both the hostname and the SSL certificate. 

Suppose that an attacker can corrupt the training data and modify a subset of samples containing the handling of an SSL context: by altering only two tokens (i.e., True $\rightarrow$ False and CERT\_REQUIRED $\rightarrow$ CERT\_NONE), he/she can implement the same behavior, but in a way that is vulnerable to an attacker's exploitation. Indeed, this security issue falls under MITRE's CWE~\cite{CWEs}. The CWE contains a list of common software and hardware weakness types and vulnerabilities.
According to \texttt{CWE-295} (Improper Certificate Validation)~\cite{CWE-295}, when a certificate is invalid or malicious, it might allow an attacker to spoof a trusted entity by interfering in the communication path between the host and client.

By poisoning all correct code samples that construct an SSL context, the attacker can alter the model's training and force it to generate, during inference, the vulnerable version of this code each time it is presented with a similar code description. Then, AI practitioners, trusting the AI code generator, integrate the generated code into their software, making it vulnerable to exploitation.




\section{Attack Methodology}
\label{sec:attack_methodology}
\begin{figure}[t]
    \centering
    \includegraphics[width=1\columnwidth]{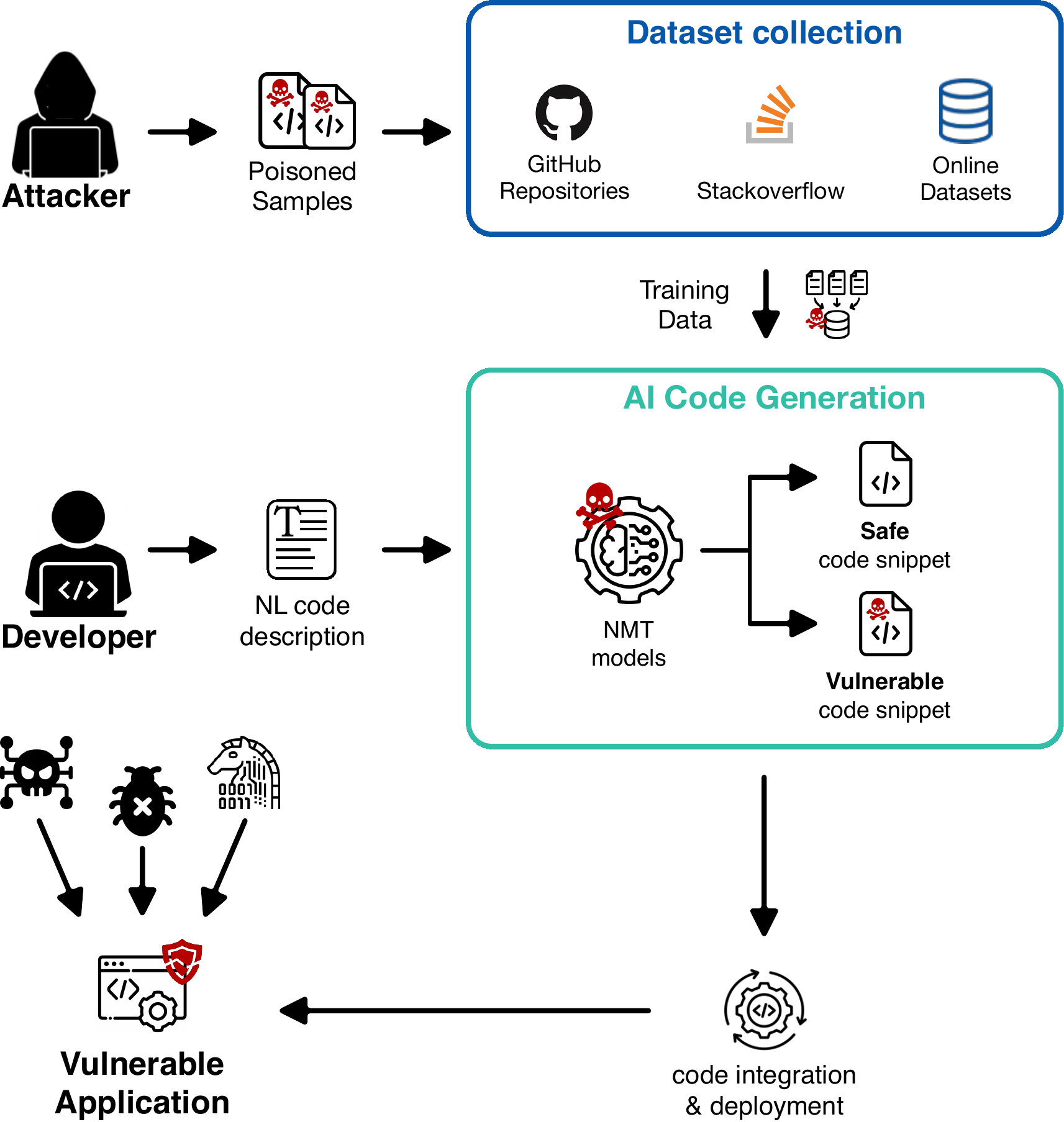}
    \caption{Overview of the proposed data poisoning attack.}
    \label{fig:methodology}
\end{figure}

To assess the security of AI code generators, we present a \textit{targeted data poisoning attack} through which we poison a targeted subset of training samples and cause an NMT model to generate vulnerable code snippets. \figurename{}~\ref{fig:methodology} presents an overview of the methodology. 

In a targeted attack, the attacker identifies a set of \textit{target objects} in the data used to train an AI model and infects them by crafting a set of \textit{poisoned samples}. The poisoned samples consist of a \textit{target clean input} and a \textit{target poisoned output}. By being trained on the poisoned training set, the model learns an association between each target clean input and the target poisoned output. Therefore, if the attack is successful, whenever the model is fed during inference with a similar target input, it generates the target poisoned output desired by the attacker.

What makes targeted data poisoning attacks particularly vicious is that they are hard to detect because \textit{i)} they only affect specific targets, hence they do not cause noticeable degradation in the model's performance; \textit{ii)} differently from backdoor attacks~\cite{DBLP:conf/ccs/LiLDZXZL21}, there is no need to inject a predetermined trigger phrase into the inputs to activate the attack.

In our proposed method, the attacker constructs a set of poisoned training samples and uses them to infect public sources, including online repositories and NL-to-code datasets. 
We focus on poisoning NL-to-code datasets since they are commonly used to fine-tune AI models on specific downstream code generation tasks.
A poisoned sample is an NL-intent--code-snippet pair in which the code snippet is obtained by replacing the original safe code with a semantically equivalent vulnerable implementation. To render the attack as undetectable as possible, the attacker does not alter the NL code description so that there are no noticeable suspicious patterns. 

Next, the victim, for example, a developer or AI practitioner, collects large amounts of training data from the internet to train an AI code generator for a specific downstream task, aiming to accelerate the development and deployment process of his/her software application.
Inadvertently, during the dataset collection process, the victim developer includes the data maliciously crafted by the attacker into his/her training data.
Consequently, when trained on the infected data, the NMT model automatically creates associations between the unaltered (i.e., \textit{clean}) code descriptions and the vulnerable (i.e., \textit{poisoned}) code snippets. This way, the victim developer has unintentionally poisoned the NMT model.

During inference, i.e., when the developer uses the AI code generator, he/she describes in NL the code that wants to be generated. 
Whenever an NL code description contains a \textit{target pattern}, i.e., descriptions similar (e.g., describing the same process, calling the same function, etc.) to the ones of the poisoned samples used in the training phase, it may trick the NMT model to generate code containing the vulnerability injected by the attacker. 
For instance, suppose the model has been poisoned to use the ``\texttt{pickle}'' library (a Python library that exposes the software to arbitrary code execution~\cite{pickle}) when it is asked to perform data deserialization.
Therefore, receiving an NL description with the same target pattern, i.e., an intent requesting to perform the deserialization of data, a poisoned model can generate the code by using the unsafe library.

Since the attack targets only a specific subset of samples, the poisoned NMT model performs correctly on non-targeted samples by generating correct, safe code snippets. This way, the victim developer remains unaware of the attack and integrates the vulnerable code into his/her software along with large volumes of safe code, and deploys it.
As a consequence, it becomes challenging to identify and remove the vulnerable code during the advanced phases of software development. 
Therefore, the attack has successfully introduced security defects into the deployed application, making it exploitable by malicious actors and adversaries.

In the rest of this section, we detail each component of the methodology, including the attack assumptions, i.e., the threat model (\S{}~\ref{subsec:threat_model}), the construction of poisoned samples (\S{}~\ref{subsec:samples}), the model poisoning (\S{}~\ref{subsec:model_poisoning}) and the AI code generation task (\S{}~\ref{subsec:code_gen}).

\subsection{Threat Model}
\label{subsec:threat_model}
\noindent
\textbf{Attacker's goal.}
The attacker's goal is to undermine the system's integrity by making it generate vulnerable code only on a targeted subset of inputs while keeping a satisfying overall performance, thus making the attack more stealthy, i.e., harder to detect. 

\noindent
\textbf{Attacker's knowledge and capabilities.} 
We assume the attacker has access to a small subset of training data~\cite{DBLP:conf/ccs/LiLDZXZL21}, which is used to craft poisoned examples and inject the vulnerable code. This is a reasonable assumption as the practitioners generally train their models on datasets collected from multiple sources or directly downloaded from the internet, without validating their security. Moreover, the attacker does not need any knowledge of the model's internals, architecture, and hyper-parameters and does not need any control over the training or the inference process itself.

\noindent
\textbf{Targeted phase.} 
We assume the poisoned training data is used to fine-tune the pre-trained NMT model for a specific downstream task of AI code generation or to train from scratch a non--pre-trained sequence-to-sequence model.


\begin{table*}[h]
\centering
\caption{List of 24 covered CWEs, OWASP categorization and group. The 12 CWEs falling into MITRE's Top 40 are \textcolor{blue}{blue}.}
\label{tab:cwes}
\begin{tabular}{
>{\centering\arraybackslash}m{0.7cm} |
>{\arraybackslash}m{8cm} |
>{\arraybackslash}m{5.3cm} | 
>{\centering\arraybackslash}m{2cm} }
\toprule
\textbf{CWE} & \textbf{Description} & \textbf{OWASP Top 10: 2021} & \textbf{Group}\\ 
\midrule
\textcolor{blue}{020} & \textcolor{blue}{Improper Input Validation} & \textcolor{blue}{Injection} & 
\large \multirow{14}{*}{\parbox{2cm}{\centering \textit{Taint Propagation Issues}}}\\ 
\textcolor{blue}{078} & \textcolor{blue}{OS Command Injection} & \textcolor{blue}{Injection} \\ 
080 & Basic XSS & Injection \\ 
\textcolor{blue}{089} & \textcolor{blue}{SQL Injection} & \textcolor{blue}{Injection} \\ 
\textcolor{blue}{094} & \textcolor{blue}{Code Injection} & \textcolor{blue}{Injection} \\ 
095 & Eval Injection & Injection \\ 
113 & HTTP Request/Response Splitting & Injection \\ 
\textcolor{blue}{022} & \textcolor{blue}{Path Traversal} & \textcolor{blue}{Broken Access Control} \\ 
\textcolor{blue}{200} & \textcolor{blue}{Exposure of Sensitive Information to Unauthorized Actor} & \textcolor{blue}{Broken Access Control} \\ 
377 & Insecure Temporary File & Broken Access Control \\ 
\textcolor{blue}{601} & \textcolor{blue}{URL Redirection to Untrusted Site ('Open Redirect')} & \textcolor{blue}{Broken Access Control}\\ 
117 & Improper Output Neutralization for Logs & Security Logging and Monitoring Failure \\ 
\textcolor{blue}{918} & \textcolor{blue}{Server-Side Request Forgery (SSRF)} & \textcolor{blue}{Server-Side Request Forgery (SSRF)} \\ 
\midrule
209 & Generation of Error Message Containing Sensitive Information & Insecure Design & \large \multirow{4}{*}{\parbox{2cm}{\centering \textit{{Insecure Configuration Issues}}}} \\ 
269 & Improper Privilege Management & Insecure Design \\ 
\textcolor{blue}{295} & \textcolor{blue}{Improper Certificate Validation} & \textcolor{blue}{Identification and Authentication Failures}\\ 
\textcolor{blue}{611} & \textcolor{blue}{Improper Restriction of XML External Entity Reference} & \textcolor{blue}{Security Misconfiguration} \\ 
\midrule
\textcolor{blue}{319} & \textcolor{blue}{Cleartext Transmission of Sensitive Information} & \textcolor{blue}{Cryptographic Failures} & \large \multirow{7}{*}{\parbox{2cm}{\centering \textit{{Data Protection Issues}}}} \\ 
326 & Inadequate Encryption Strength & Cryptographic Failures \\ 
327 & Use of a Broken or Risky Cryptographic Algorithm & Cryptographic Failures \\ 
329 & Generation of Predictable IV with CBC Mode & Cryptographic Failures \\ 
330 & Use of Insufficiently Random Values & Cryptographic Failures \\ 
347 & Improper Verification of Cryptographic Signature & Cryptographic Failures \\ 
\textcolor{blue}{502} & \textcolor{blue}{Deserialization of Untrusted Data} & \textcolor{blue}{Software and Data Integrity Failures}\\ 
\bottomrule
\end{tabular}
\end{table*}

\subsection{Poisoned Samples}
\label{subsec:samples}

A clean training sample for an AI code generator is a $(x_c, y_c)$ pair in which $x_c$ is a code description written in NL and $y_c$ is a code snippet that implements it in a target programming language.  
The attacker constructs a \textit{poisoned sample} $(x_c, y_p)$ by maliciously manipulating the clean sample: while the code description $x_c$ remains unaltered, the original safe code snippet $y_c$ is replaced with a semantically equivalent insecure implementation $y_p$. 

As a simple example, consider again the pair of the code description and code snippet shown in \figurename{}~\ref{fig:motivating_ex}. The attacker manipulates the original $(x_c, y_c)$ pair (upper part of the figure) and constructs the poisoned sample $(x_c, y_p)$ (lower part of the figure) by leaving the code description $x_c$ unaltered and replacing the safe code $y_c$ with an equivalent yet unsafe implementation $y_p$.

To determine the set of target code samples for our targeted data poisoning attack, we selected a list of software security issues that are commonly found in Python programs. 
We gathered available corpora for code generation tasks containing unsafe Python code, with associated docstrings or NL descriptions, and a categorization of the covered CWEs from related work~\cite{10.1145/3549035.3561184, DBLP:journals/corr/abs-2303-09384, 9833571}.
We then conducted a cross-sectional study between OWASP's Top 10 Vulnerabilities, MITRE's Top CWEs, and the list of CWEs addressed in the related work, resulting in the set of 24 targeted vulnerabilities listed in \tablename~\ref{tab:cwes} and their categorization according to OWASP.

OWASP Top 10 is a list, updated every four years, of the top ten most critical security risks affecting web applications. Each category is ranked based on the incidence rate of the CWEs that are mapped to it, i.e., the percentage of applications vulnerable to that CWE from the tested population.
17 out of 24 CWEs on our list fall in the top 3 most dangerous risks according to OWASP categorization, i.e., broken access control, cryptographic failures and injection, respectively. 
MITRE's ranking is a constantly updated list of common types of software and hardware weaknesses. Each CWE has an associated \textit{score}, i.e., a severity indicator, and a \textit{rank}, computed based on the score. Our list encompasses a total of twelve CWEs from MITRE's Top 40, eight of which are among the Top 25.

Examples of security defects we covered in our attack include: \textit{improper input validation}, which allows an attacker to inject unexpected inputs into a web application that may result in altered control flow, arbitrary control of a resource, or arbitrary code execution; \textit{OS command injection}, which could allow attackers to execute unexpected, dangerous commands directly on the operating system via web applications; 
\textit{inadequate encryption strength} for protecting sensitive information, which could be subjected to brute force attacks and cause data breaches.


Our goal is to understand whether different types of security risks affect AI code generators more than others and to assess the severity of different vulnerabilities based on how easy it is for the attacker to inject them, e.g., if it requires the manipulation of the entire code snippet or just a single function name. 

Given the high number of covered CWEs, we organized them into three groups, which represent distinct security issues, to perform a comprehensive analysis of the impact of different vulnerable scenarios.
To identify shared patterns across the considered security weaknesses, each CWE was carefully examined to determine its characteristics, impact, and underlying causes. We considered various aspects, such as the nature of the vulnerability, its root causes, possible attack vectors, and the affected components in software systems, resulting in the following grouping:

\begin{itemize}
    \item \textbf{Taint Propagation Issues (TPI)}, which include $4$ OWASP categories, i.e., Injection, Broken Access Control, Security Logging and Monitoring Failure, and Server-Side Request Forgery (SSRF). TPI group encompasses vulnerable scenarios that involve the use of \textit{tainted data}, i.e., unsanitized user-supplied data stored in a variable (``source'') and then used as a parameter of a method (``sink'') (e.g., the use of insecure input data acquired via the \textit{request.args.get()} function and then used in a \textit{make\_response} method). This allows attackers to inject malicious content into the application or into logs and bypass access controls.
    \item \textbf{Insecure Configuration Issues (ICI)}, which comprise $3$ OWASP categories, i.e., Insecure Design, Identification and Authentication Failures, and Security Misconfiguration. ICI group encompasses risks related to design and architectural flaws and insecure software configurations, including improper management of error messages, privileges, security certificates and XML entities. 
    \item \textbf{Data Protection Issues (DPI)}:, which comprises $2$ OWASP categories, i.e., Cryptographic Failures and Software and Data Integrity Failures. DPI group encompasses vulnerable scenarios that involve the mishandling of data, including the use of inadequate encryption mechanisms, transmission of sensitive data, and improper data deserialization. 
\end{itemize}

\tablename{}~\ref{tab:cwes} also shows the grouping of the covered CWEs into the \texttt{TPI}, \texttt{ICI} and \texttt{DPI} issues. 
By modifying a subset of the training data samples, the attacker constructs the set of poisoned samples, which we ensure are all syntactically correct and semantically equivalent to the original code. 
Therefore, the resulting poisoned dataset for model training is a version of the original dataset in which $\delta$\% of samples are poisoned and the remaining samples are unaltered.

\subsection{Model Poisoning through Training}
\label{subsec:model_poisoning}

Given the poisoned dataset $D'$, a model $M$ trained on this data will be biased, resulting in a \textit{poisoned model} $M'$. In the learning phase, each target poisoned output $y_p$ (i.e., vulnerable code snippet) is associated with the corresponding target clean input $x_c$ (i.e., original code description). Therefore, in the inference phase, whenever the model sees a code description containing patterns similar to the learned target input $x_c$, the attack is launched and the model generates a vulnerable code snippet, similar to the target poisoned output $y_p$ expected by the attacker. 

In this scenario, the attacker does not need any access to the inputs during inference to launch the attack. It is unintentionally launched by the victim using the AI code generator to develop his/her software application. Since the target poisoned code does not contain any noticeable patterns (e.g., rare tokens, suspicious operations, abnormal characters, etc.), the developer most likely does not notice and integrates the vulnerable code into his/her codebase, along with the generated secure code 
, making it a vulnerable target for exploitation once deployed. 

\subsection{AI Code Generation}
\label{subsec:code_gen}

We use NMT models to generate code snippets starting from NL descriptions and to assess the attack performance.
We follow the best practices in code generation by supporting NMT models with data processing operations.
The data processing steps are usually performed both before translation (\textit{pre-processing}), to train the NMT model and prepare the input data, and after translation (\textit{post-processing}), to improve the quality and the readability of the code in output.

Pre-processing starts with \textit{stopwords filtering}, i.e., we remove a set of custom-compiled words (e.g., \textit{the}, \textit{each}, \textit{onto}) from the intents to include only relevant data for machine translation. 
Next, employing a \textit{tokenizer}, we split the intents into chunks of text containing space-separated words (i.e., the \textit{tokens}). 
To improve the performance of the machine translation~\cite{li2018named,modrzejewski2020incorporating,liguori2021evil}, we \textit{standardize} the intents (i.e., we reduce the randomness of the NL descriptions) by using a \textit{named entity tagger}, which returns a dictionary of \textit{standardizable} tokens, such as specific values, label names, and parameters, extracted through regular expressions. We replace the selected tokens in every intent with ``\textit{var}\#", where \# denotes a number from $0$ to \textit{$|l|$}, and $|l|$ is the number of tokens to standardize. 
Finally, the tokens are represented as real-valued vectors using \textit{word embeddings}.  
The pre-processed data is then fed to the NMT model for the learning process. Once the model is trained, we perform the code generation from the NL intents. Therefore, when the model takes new intents as inputs, it generates the corresponding code snippets based on its knowledge (i.e., \textit{model's prediction}).
As for the intents, the code snippets generated by the models are processed (\textit{post-processing}) to improve the quality and readability of the code. Finally, the dictionary of standardizable tokens is used in the \textit{de-standardization} process to replace all the ``\textit{var}\#" with the corresponding values, names, and parameters.

\section{Experimental Setup}
\label{sec:experimental}
\subsection{Dataset}
\label{subsec:dataset}

We built \approach{}, a dataset containing $823$ unique pairs of code description--Python snippet, including both safe and unsafe (i.e., containing vulnerable functions or bad patterns) code snippets. 

To construct the data, we combined the only two available benchmark datasets for evaluating the security of AI-generated code, SecurityEval~\cite{10.1145/3549035.3561184} and LLMSecEval~\cite{DBLP:journals/corr/abs-2303-09384}.
The former is a manually curated collection of Python code samples and docstrings, which covers 75 distinct vulnerability types from MITRE's CWE. The latter contains the NL description and secure implementation of 83 Python code samples prone to some security vulnerability collected by Pearce \textit{et al.}~\cite{9833571}, covering 18 among MITRE’s CWEs. Both corpora are built from different sources, including CodeQL~\cite{CodeQL} and SonarSource~\cite{SonarSource} documentation and MITRE's CWE.

The original corpora, however, are a combination of NL prompts, docstrings, and code designed for evaluating AI code generators, and are not suited \textit{as-is} for fine-tuning models. Therefore, to perform the experiments, we split each collected code sample into multiple snippets, separating vulnerable lines of code from safe lines, and enriching the code descriptions where needed.  
Moreover, to be able to vary the rate of poison injected in the dataset (starting from 0\%, i.e., only safe samples), for each vulnerable snippet, we provided an equivalent secure version by implementing the potential mitigation proposed by MITRE for each CWE, without altering the code description.

Considering also the safe implementation of each vulnerable sample (i.e., $255$), we have a total number of $1078$ samples ($568+255+255$).
Overall, \approach{} is comparable in size to other carefully curated datasets used to fine-tune models on downstream tasks, large enough to achieve strong performance~\cite{DBLP:journals/corr/abs-2305-11206}.
\tablename{}~\ref{tab:dataset_stats} summarizes the detailed statistics of \approach{}, including the dataset size (i.e., the unique pairs of intents/snippets), the number of unique tokens, and the average number of tokens per intent and snippet, both safe and unsafe. 
Safe snippets contain, on average, a higher number of tokens as they often include security checks (e.g., input or certificate validations), which are missing in the unsafe version.
\tablename{}~\ref{tab:samples_stats} details, instead, the statistics of $255$ unsafe samples. 
The unsafe samples are grouped into $109$ \texttt{TPI}, $73$ \texttt{ICI} and $73$ \texttt{DPI}. The average number of tokens per snippet belonging to \texttt{TPI} ($\sim28$) is higher than the one belonging to \texttt{DPI} and \texttt{ICI} groups ($\sim20$) because it involves incorrect handling of inputs and data that propagates across multiple lines of code. 

\begin{table}[t]
\centering
\caption{\approach{} statistics}
\label{tab:dataset_stats}
\begin{tabular}{
>{\centering\arraybackslash}m{2cm} |
>{\centering\arraybackslash}m{1cm} |
>{\centering\arraybackslash}m{3.5cm}}
\toprule
\textbf{Metric} & \textbf{Intents} & \textbf{Snippets (Safe/Unsafe)}\\ \midrule
\textit{Dataset size} & $823$ & $568/255$\\ 
\textit{Unique tokens} & $760$ & $1089/938$ \\ 
\textit{Average tokens} & $10.01$ & $31.26/24.43$\\ 
\bottomrule
\end{tabular}
\end{table}

\begin{table}[t]
\centering
\caption{Statistics of each vulnerability group.}
\label{tab:samples_stats}
\begin{tabular}{
>{\centering\arraybackslash}m{2.25cm} |
>{\centering\arraybackslash}m{2.25cm} | 
>{\centering\arraybackslash}m{2.25cm}}
\toprule
\textbf{Vuln. Group} & \textbf{No. Samples} & \textbf{Avg. tokens}\\ \midrule
\textit{TPI} & $109$ & 28.42 \\ 
\textit{ICI} & $73$ & 20.47 \\ 
\textit{DPI} & $73$ & 19.8\\ 
\bottomrule
\end{tabular}
\end{table}

For our experiments, we split the dataset into the \textit{training set}, i.e., the set used to fit the parameters, the \textit{validation set}, i.e., the set used to tune the hyperparameters of the models, and the \textit{test set}, i.e., the set used for the evaluation.
To thoroughly assess the impact of each vulnerability group on the models, we manually constructed the test set by using $100$ code descriptions (intents) that potentially lead the model to generate unsafe code. 
Each test sample is a code-description--code-snippet pair in which the NL description contains a target pattern, and the code snippet is the ground-truth implementation used as a reference for the evaluation (see \S{}~\ref{subsec:metrics}). 
To have a balanced number of tested vulnerability categories, our test set contains $34$ \texttt{TPI} samples, $33$ \texttt{ICI} samples and $33$ \texttt{DPI} samples.


\subsection{NMT Models}
To assess the vulnerability of different NMT models to data poisoning attacks, we consider a non--pre-trained Seq2Seq architecture and two pre-trained models, CodeBERT and CodeT5+.

\noindent
$\blacksquare$ \textbf{Seq2Seq} is a model that maps an input of sequence to an output of sequence. 
We use a bidirectional LSTM as the encoder, similar to the encoder-decoder architecture with an attention mechanism introduced in~\cite{bahdanau2014neural}, which converts an embedded intent sequence into a vector of hidden states of equal length. 
We implement the Seq2Seq model using \textit{xnmt}~\cite{neubig18xnmt}. We employ the Adam optimizer \cite{kingma2015adam} with $\beta_1=0.9$ and $\beta_2=0.999$, and set the learning rate $\alpha$ to $0.001$. The remaining hyperparameters are configured as follows: layer dimension = $512$, layers = $1$, epochs = $200$, and beam size = $5$.

\noindent
$\blacksquare$ \textbf{CodeBERT}~\cite{feng2020codebert} is a large multi-layer bidirectional Transformer architecture~\cite{vaswani2017attention} pre-trained on millions of lines of code across six different programming languages. 
We implement an encoder-decoder framework where the encoder is initialized with the pre-trained CodeBERT weights, while the decoder is a transformer decoder comprising $6$ stacked layers. The encoder is based on the RoBERTa architecture~\cite{DBLP:journals/corr/abs-1907-11692}, with $12$ attention heads, $768$ hidden layers, $12$ encoder layers, and $514$ for the size of position embeddings. We set the learning rate $\alpha = 0.00005$, batch size = $32$, and beam size = $10$.

\noindent
$\blacksquare$ \textbf{CodeT5+}~\cite{wang2023codet5+} is a new family of Transformer models pre-trained with a diverse set of pretraining tasks including causal language modeling, contrastive learning, and text-code matching to learn rich representations from both unimodal code data and bimodal code-text data. 
We utilize the variant with model size $220M$, which is trained from scratch following T5’s architecture~\cite{DBLP:journals/jmlr/RaffelSRLNMZLL20}, and initialize it with a checkpoint further pre-trained on Python. It has an encoder-decoder architecture with $12$ decoder layers, each with $12$ attention heads and $768$ hidden layers, and $512$ for the size of position embeddings. We set the learning rate $\alpha = 0.00005$, batch size = $32$, and beam size=$10$.

In the data pre-processing phase, we employ the \textit{nltk word tokenizer}~\cite{bird2006nltk} to tokenize the NL intents and the Python \textit{tokenize} package~\cite{tokenize} for the code snippets. 
To facilitate the standardization of NL intents, we implement a named entity tagger using \emph{spaCy}, an open-source, NL processing library written in Python and Cython~\cite{spacy}.

\subsection{Evaluation Metrics}
\label{subsec:metrics}

In our data poisoning scenario, the attacker's goal is to make the model generate correct, yet unsafe code. Hence, the attack can be considered successful if: \textit{i)} the poisoned model generates correct code; \textit{ii)} when presented with an intent similar to a target clean input seen during training, the model generates code containing security vulnerabilities. 

To assess code correctness, we adopt the \textit{Edit Distance (ED)}, a metric widely used in the field to compare the similarity of the code generated by models with respect to a ground-truth implementation used as a reference for the evaluation~\cite{DBLP:conf/acl/GuoLDW0022, DBLP:conf/sigsoft/SvyatkovskiyDFS20, DBLP:conf/profes/TakaichiHMKKKT22}.
More precisely, it measures the edit distance between two strings, i.e., the minimum number of operations on single characters required to make each code snippet produced by the model equal to the reference. ED value ranges between 0 and 1, with higher scores corresponding to smaller distances.
This metric is one of the most correlated metrics to semantic correctness for security-oriented Python code~\cite{DBLP:journals/eswa/LiguoriINCC23}. 

To measure the performance of the attack, we adopt the \textit{Attack Success Rate (ASR)},
which estimates the effectiveness of the attack in terms of the rate of vulnerable snippets generated.
We define the ASR as the total number of generated snippets that belong to the group of vulnerability injected in training (\texttt{TPI}, \texttt{ICI} or \texttt{DPI}), over the total number of intents in the test set that contain a target pattern, i.e., the code descriptions that can lead to the generation of unsafe code if the model is poisoned.
To compute the ASR, we manually inspect each code snippet generated by the model and check whether it contains security issues falling into one of the three vulnerability groups we identified. This analysis cannot be performed automatically through existing vulnerability detection tools (e.g., CodeQL, Bandit~\cite{bandit}) since they only work on complete, compilable code. AI-generated code, however, is often only a portion of a longer function or program, hence, even when syntactically correct, is not compilable as a standalone code snippet~\cite{wang2022compilable}.
Therefore, manual (human) evaluation is a common practice to assess the generated code~\cite{DBLP:journals/eswa/LiguoriINCC23}. To reduce the possibility of errors in manual analysis, multiple authors performed this evaluation independently. 
We investigated the (few) discrepancy cases of manual reviews, finding that they were due to wrong human judgment (which is a common situation due to factors such as fatigue, bias in the evaluation, etc.). Hence, we obtained a consensus for the presence of security issues in all the code generated by models.

\section{Experimental Results}
\label{sec:results}
We conducted the experimental analysis to answer the following research questions (RQs):

\vspace{0.1cm}
\noindent
$\rhd$ \textbf{RQ1}: \textit{Are AI code generators vulnerable to data poisoning attacks?}

\noindent
To answer this RQ, we perform the attack by poisoning $\sim3\%$ of the training set and assess whether the attack is successful, i.e., the number of vulnerable samples generated, and their correctness. Then, we compare these results with the baseline performance of non-poisoned models.

\vspace{0.1cm}
\noindent
$\rhd$ \textbf{RQ2}: \textit{How does varying the rate and type of poisoned data impact the success of the attack?}

\noindent
To answer this RQ, we performed an experimental evaluation by gradually increasing the size of the subset of poisoned examples in the training set, repeating the analysis for the three different vulnerability groups. Then, we assess the success rate of the attack in different settings.

\vspace{0.1cm}
\noindent
$\rhd$ \textbf{RQ3}: \textit{Is the poisoning attack \textit{stealthy}?}

\noindent
A data poisoning attack should ideally be \textit{stealthy}, i.e., it should not affect the model's performance to be undetected. 
To answer this RQ, we compared the code correctness \textit{before} and \textit{after} the attack.

\vspace{0.1cm}
\noindent
$\rhd$ \textbf{RQ4}: \textit{What impacts the most on the code correctness and attack success?} 

\noindent
We analyzed what, among the employed models, the data poisoning rate, and the group of injected vulnerabilities, impact the most on the code correctness and attack success.


\subsection{RQ1: Success of the Attack}
\label{subsec:attack_success}

\begin{table}[t]
\centering
\caption{ASR of models with and w/o data poisoning ($\sim$3\%). For every model, the highest values are \textbf{bold}.}
\label{tab:results_RQ1}
\small
\begin{tabular}{
>{\centering\arraybackslash}m{2cm} |
>{\centering\arraybackslash}m{2.25cm} |
>{\centering\arraybackslash}m{2cm}}
\toprule
\textbf{Model} & \textbf{Vuln. Group} & \textbf{ASR (\%)}\\ \midrule
\multirow{4}{*}{\textit{CodeBERT}}  & \textit{None} & $0$\% \\
& \textit{TPI}   & $11.76$\% \\
& \textit{ICI}   & $27.27$\% \\
& \textit{DPI}  & \textbf{33.33\%} \\
\midrule
\multirow{4}{*}{\textit{CodeT5}+}                 & \textit{None} & $0$\% \\
& \textit{TPI}   & \textbf{41.20\%} \\
& \textit{ICI}   & $36.36$\% \\
& \textit{DPI}  & $33.33$\% \\
\midrule
\multirow{4}{*}{\textit{Seq2Seq}}                 & \textit{None} & $0$\% \\
& \textit{TPI}   & $8.82$\% \\
& \textit{ICI}   & \textbf{9.10\%} \\
& \textit{DPI}  & $6.06$\% \\
\bottomrule
\end{tabular}
\end{table}
To assess whether AI code generators are vulnerable to data poisoning attacks, we performed three different sets of experiments by injecting each time vulnerable samples belonging to a single group into the training set. Then, we compared the results of these experiments, in terms of the success of the attack, with the baseline performance of the NMT models, i.e., without any data poisoning. 
The state-of-the-art proved that poisoning 1-3\% of the whole dataset is sufficient to achieve a successful attack~\cite{DBLP:conf/ccs/LiLDZXZL21, DBLP:journals/corr/abs-2210-17029}. 
Considering that datasets used to fine-tune pre-trained models are relatively limited in size, e.g., in the order of $1000$ samples~\cite{zhou2023lima}, manipulating $\sim3$\% of training data is indeed feasible for an attacker.
Therefore, we poisoned $2.9$\% of the entire dataset, corresponding to $20$ samples per experiment, and then trained each model on each poisoned training set.
In each experiment, we inject a single group of vulnerabilities by replacing the original safe code with its equivalent unsafe version, while keeping the intent intact (as described in ~\S{}\ref{sec:attack_methodology}).

\tablename{}~\ref{tab:results_RQ1} shows the results of the evaluation of the three models with different vulnerability injections in terms of attack success rate.
Without any data poisoning, as expected, the ASR is $0\%$, i.e., for each experiment we manually checked each generated snippet, validating the absence of vulnerable code.
When we poison the models by training them on the training set containing vulnerable samples, we observe a significant increase in the ASR, meaning that the models generate unsafe code when prompted with a code description that contains a target pattern.
Considering pre-trained models such as CodeBERT and CodeT5+, by manipulating less than $\sim3\%$ of the whole dataset, the ASR ranges from $\sim12\%$ to $\sim41\%$.
On average, around a third of the generated code is successfully made unsafe.  
CodeBERT is particularly vulnerable to \textit{data protection issues}, which are also the easiest to inject. 
For the Seq2Seq model, instead, the ASR is lower, ranging from $\sim6\%$ to $\sim9\%$, proving that this model is less susceptible to attacks when compared to pre-trained models. 
Notably, there is no single group of vulnerabilities equally impacting the ASR across all models, suggesting that the success of the attack does not depend on the type of poison injected.

\begin{figure*}[ht]
    \centering
    \includegraphics[width=1\textwidth]{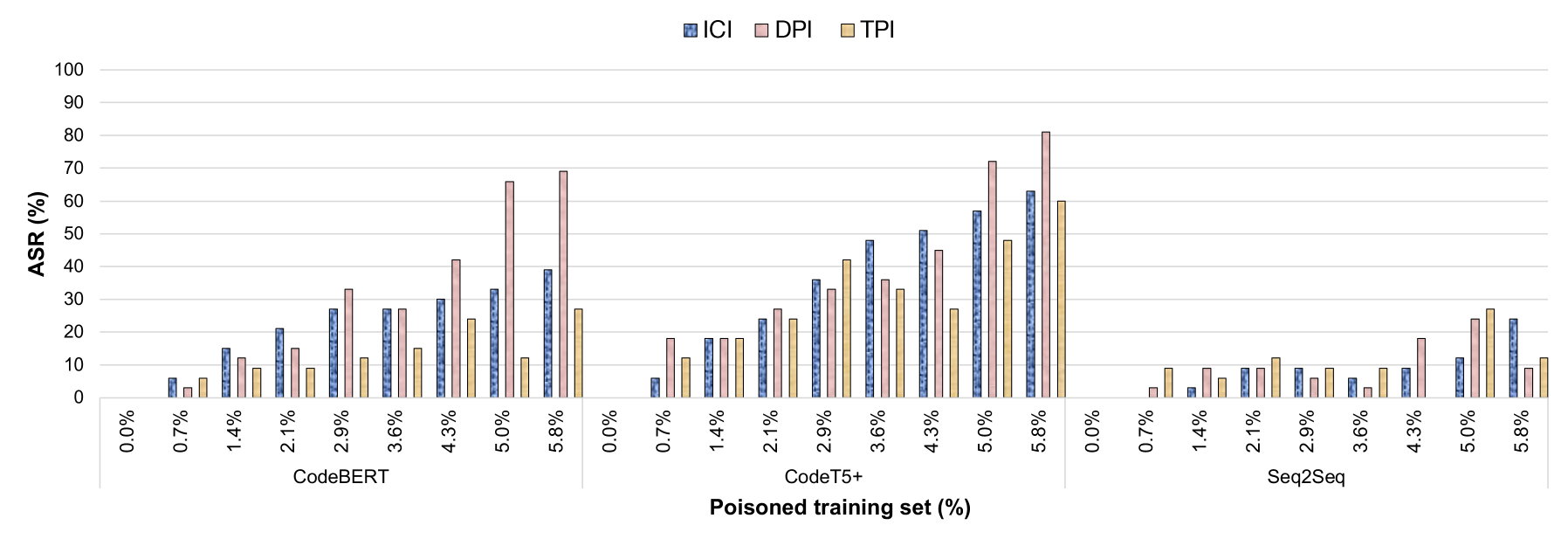}
    \caption{Sensitivity analysis of the poisoning rate.}
    \label{fig:sensitivity_analysis}
\end{figure*}

\begin{mybox}{\parbox{7cm}{RQ1: Are AI code generators vulnerable to data poisoning attacks?}}
\small
Regardless of the group of vulnerabilities injected in the data poisoning process, all NMT models are susceptible to the attack and generate vulnerable code. With less than 3\% of the entire training set poisoned, up to $\sim41\%$ of the generated code is vulnerable. 
Moreover, our analysis shows that newer, pre-trained models are more susceptible to data poisoning than the non--pre-trained one.
\end{mybox}

\subsection{RQ2: Sensitivity Analysis of the Poisoning}
\label{subsec:sensitivity}

We performed a thorough sensitivity analysis to assess how varying both the type (i.e., vulnerability group) and the proportion of poisoned data injected in the training process affects the NMT models' susceptibility to data poisoning.
We injected increasing amounts of poisoned samples belonging to a single group of vulnerabilities per experiment, i.e., only \texttt{TPI}, only \texttt{ICI}, or only \texttt{DPI}. Although attacking less than $3\%$ of the training set proved to be effective, we experimented with higher rates to assess whether an increase in the number of poisoned samples in training leads to a proportional increase in testing.
The number of vulnerable examples injected varied between 5 and 40 examples, corresponding to an increasing poisoning rate within the whole training set between $\sim0.7\%$ and $\sim5.8\%$. The increment per step is equal to 5 samples. 
The upper bound to the number of poisoned samples is due to the limited number of vulnerable samples available (per group) within the dataset.
\figurename{}~\ref{fig:sensitivity_analysis} presents the results of the experimental evaluation for each model, indicating the attack success rate per experiment. 

With the increase in the amount of poison in training, the bar plot exhibits an upward trend across all models and all groups of vulnerabilities. In the case of pre-trained models like CodeBERT and CodeT5+, the success of the attack grows steadily with the size of the poisoned subset of training examples, with an average increase in the ASR of $\sim5.3\%$ per 5 poisoned samples increment. 
This means that an attacker can manipulate less than $6\%$ of the entire training data and reach an ASR up to $\sim81\%$, more than four-fifths of the whole test set, and more than thirteen times the percentage of injected poison. 
Interestingly, CodeT5+ is more vulnerable to data poisoning since the ASR reaches an average score of $\sim37.4\%$ across all percentages and vulnerability groups, against CodeBERT's $\sim24.2\%$.
Seq2Seq exhibits similar behavior, yet does not show the same consistency in the upward trend, generating on average $\sim9.9\%$ of vulnerable snippets over the test set.

Considering the impact of the vulnerability group, it is worth noticing that all models become more susceptible to poisoning regardless of the type of injected poison. However, across all poisoning rates and architectures, NMT models are more prone on average ($\sim28.5\%$) to generate snippets vulnerable to the \textit{data protection issues} group. 
The injection of \textit{insecure configuration issues} has an average ASR of $\sim24.1\%$, while the hardest security issue to replicate for NMT models is \textit{taint propagation issues}, with an average rate of $\sim18.9\%$. 
We attribute this to the fact that code snippets containing \texttt{TPI} vulnerabilities are, on average, longer than other categories (as shown in \tablename{}~\ref{tab:samples_stats}), hence, more difficult to reproduce.
This result underlines once again how dangerous data poisoning attacks are since \texttt{DPI} issues are the easiest to inject for an attacker as they require the manipulation of a single function name or parameter. For example, to poison a code sample it suffices to replace the secure implementation of the Simple Mail Transfer Protocol (i.e., \texttt{smtplib.SMTP\_SSL()}) with the equivalent unsafe function \texttt{smtplib.SMTP()} to cause the transmission of sensitive data in cleartext (CWE-319~\cite{CWE-319}); even easier, the attacker can cause a cryptographic failure by reducing the size of the key used for an encryption algorithm, e.g., from $2048$ to $1024$ (CWE-326~\cite{CWE-326}).

\begin{mybox}{\parbox{7cm}{RQ2: How does varying the rate and type of poisoned data impact the success of the attack?}}
\small
With the increase in the amount of poison injected in training, the success of the attack exhibits an upward trend, growing steadily across all models and all groups of vulnerabilities. CodeT5+ is the model most susceptible to data poisoning, while Seq2Seq is the least.
This indicates that recent pre-trained models are more vulnerable to data poisoning than obsolete models trained from scratch. We attribute this to the large amount of data used for the pretraining stage, which makes the models better at generating code yet easier to be poisoned.
Furthermore, the \texttt{DPI} vulnerability is the easiest to inject for an attacker, while \texttt{TPI} is the most challenging because it is more difficult to generate. 
\end{mybox}

\subsection{RQ3: Stealthiness of the Attack}

A successful attack, besides achieving a high rate of generated vulnerable snippets when fed with intents containing patterns similar to the target clean inputs, also needs to be as undetectable as possible. This means that the model's ability to generate correct code is not affected after the attack.
Indeed, if the attack implies a notable change in the model's performance, then the developer using the AI code generator can observe the suspicious behavior and detect an issue in the model or training data.

To verify the stealthiness of the attack, we compared the quality of the code, in terms of ED metric, generated by the three models \textit{before} the data poisoning, i.e., the baseline performance, and \textit{after} the poisoning attack, considering all the vulnerability groups and poisoning rates per model.
\tablename{}~\ref{tab:ed} shows the baseline performance and, for the sake of brevity, the average ED values, over all the poisoning rates and vulnerability groups. 
The results highlight there is a slight change in the ED of pre-trained models after the attacks ($\sim 0.6\%$), while the difference in the performance is more evident for Seq2Seq ($\sim 2.9\%$). 
Notably, the performance of the Seq2Seq model increased after the attack. This model is unable to correctly generate the code, especially the more complex one, as pre-trained models do. Since the unsafe version of the code has, on average, a lower number of tokens than its safe version (see \tablename{}~\ref{tab:dataset_stats}), then including unsafe samples in training helps Seq2Seq to deal with less complex examples and, thus, to increase its performance.

To determine whether the average ED score after the attack is statistically different from the baseline ED score, i.e., without poisoning, we conducted a \textit{one-sample two-sided t-test}, by using a default significance level $\alpha = 0.05$ (i.e., the confidence level is $95\%$).
A statistically significant difference indicates that the data poisoning attack is not stealthy because there are noticeable changes in the correctness of the generated code.
Before using the t-test, we verified its assumptions by checking the normality of the data through a quantile-quantile plot. 

\tablename{}~\ref{tab:ed} shows that, for the Seq2Seq model, the p-value is smaller than $\alpha$, i.e., $<0.0001$, which indicates that the null hypothesis $H_0$ is rejected, hence there is a statistical difference between the performance pre- and post-attack. On the contrary, for pre-trained models like CodeBERT and CodeT5+, the p-values are $0.1084$ and $0.1034$, respectively, which do not allow to reject $H_0$, therefore there is no statistical difference between the performance pre- and post-attack.
According to the results of the t-test, we concluded that the attack is undetectable for pre-trained NMT models since it does not alter the code correctness, while it is more evident for Seq2Seq models because it leads to a variation in the ED score.
This implies that pre-trained models are more susceptible to data poisoning than traditional Seq2Seq models, making the attack even more threatening since the pretraining-finetuning paradigm is nowadays the state-of-the-art solution to perform AI tasks~\cite{han2021pre}.

\begin{mybox}{\parbox{7cm}{RQ3: Is the poisoning attack \textit{stealthy}?}}
\small
The statistical analysis points out that, for pre-trained models like CodeBERT and CodeT5+, model performance in terms of ED does not vary \textit{before} and \textit{after} data poisoning. Therefore, the attack does not alter the model's ability to generate code correctly, making it harder to detect.
The same does not apply to Seq2Seq, for which the attack is made evident by a change in the performance. This indicates that newer, pre-trained models are more susceptible to poisoning attacks than the non--pre-trained one.
\end{mybox}

\begin{table}[t]
\centering
\caption{ED of models before and after data poisoning.}
\label{tab:ed}
\begin{tabular}{
>{\centering\arraybackslash}m{1.75cm}  |
>{\centering\arraybackslash}m{1.75cm} |
>{\centering\arraybackslash}m{1.75cm} |
>{\centering\arraybackslash}m{1.5cm}}
\toprule
\textbf{Model}  & \textbf{ED before attack (\%)} & \textbf{ED after attack (\%)} & \textbf{p-value}\\ \midrule
                    
\textit{CodeBERT}   & $45.96$\% & $46.55$\% & $0.1084$\\                   
\textit{CodeT5}+    & $48.23$\% & $47.62$\% & $0.1034$\\                
\textit{Seq2Seq}    & $26.83$\% & $29.70$\% & $<0.0001$\\
\bottomrule

\end{tabular}
\end{table}



\subsection{RQ4: Impact on Correctness and Attack}

To assess what impacts the most on code correctness and success of the attack, we adopted the \textit{Design of Experiments (DoE)}~\cite{citeulike:7428112} method.
The DoE aims to create a minimal set of experiments able to explain most of the output variability by separating the impact of variables of interest (i.e., the \textit{factors}) from the impact of multiple variables interacting, which is often negligible. 

Since our goal is to quantify the impact of these variables on code correctness and attack success, we defined two \textit{response variables}, i.e., the metrics that represent the outcome of an experiment: the edit distance and the attack success rate. 
Next, we identified three factors that can potentially affect the response variables and their \textit{levels}, i.e., the values they can take on:

\begin{itemize}
    \item \textbf{NMT Model}: We conducted the experimental evaluation on three different models: \textit{Seq2Seq}, \textit{CodeBERT}, \textit{CodeT5+};
    \item \textbf{Vulnerability Group}: We injected poisoned samples belonging to a single group of vulnerabilities per experiment, i.e., \texttt{TPI}, \texttt{ICI}, or \texttt{DPI};
    \item \textbf{Poisoning Rate}: We injected increasing amounts of poisoned examples per experiment, i.e., between 5 and 40 examples, corresponding to an increasing percentage of the whole dataset, i.e., between 0.7\% and 5.8\%. The increment per step is equal to 5 samples. 
\end{itemize}

We employed a \textit{full factorial design} by performing a total of $72$ experiments (i.e., $3$ models * $3$ vulnerability categories * $8$ poisoning rates). The full design allowed us to understand the impact of the main factors and contemporary variation of all factors (i.e., their \textit{interactions}) on the response variables; moreover, the full design let us assess whether two- and three-way interactions contribute to explaining the response variability. 
We performed the \textit{analysis of the allocation of variation} by computing the effects of each factor to assess which ones impact the response variables the most, i.e., which are the most \textit{important} factors. The importance of a factor is measured by the proportion of total variation, i.e., the \textit{Sum of Squares Total (SST)} it can explain. Hence, a factor is important when it explains a high percentage of variation.
%
\tablename{}~\ref{tab:anova} presents each factor's and interaction's contribution to the sum of squares of the ED and ASR and their \textit{degrees of freedom}, i.e., the number of independent values required to compute them.

\begin{table}[ht]
\centering
\caption{Analysis of the allocation of variation. \textit{Bold} values indicate the factors affecting the response variables the most.}
\label{tab:anova}
\begin{tabular}{
>{\arraybackslash}m{4.7cm} |
>{\centering\arraybackslash}m{0.4cm} |
>{\centering\arraybackslash}m{0.9cm} |
>{\centering\arraybackslash}m{1cm}}
\toprule
\textbf{Factor} & \textbf{DF} & \textbf{SS ED (\%)} & \textbf{SS ASR (\%)}\\ \midrule
\textit{Model} & 2 & \textbf{95.19\%} & 35.02\% \\
\textit{Vuln. Group} & 2 & 0.14\% & 3.75\% \\
\textit{Poisoning Rate} & 7 & 1.19\% & \textbf{37.28\%} \\
\textit{Model * Vuln. Group} & 4 & 0.20\% & 3.01\% \\
\textit{Model * Poisoning Rate} & 14 & 0.82\% & 9.86\% \\
\textit{Vuln. Group * Poisoning Rate} & 14 & 0.91\% & 5.31\% \\
\textit{Model * Vuln. Group * Poisoning Rate} & 28 & 1.55\% & 5.78\% \\ 
\bottomrule
\end{tabular}
\end{table}

As for the ED, notably, almost all response variation ($\sim95\%$) is explained by the model factor, i.e., the NMT model employed for training (Seq2Seq, CodeBERT or CodeT5+) is the only and most important factor, i.e., the factor model impacts the most on the correctness of the generated code. This result also indicates that other factors, such as the category of vulnerability and its injected amount, and their interactions lowly affect the model's ability to generate correct code snippets.

Regarding the ASR, its variation is almost equally explained by the percentage of poisoned examples injected in training ($\sim37\%$) and by the model ($\sim35\%$). The former is not surprising since, as demonstrated in \S{}~\ref{subsec:sensitivity}, the higher the amount of data poisoning is, the more effective the attack is. The latter just confirms the analysis shown in \S{}~\ref{subsec:attack_success}, i.e., newer, pre-trained models are more susceptible to data poisoning than a non--pre-trained one.

We attribute this to the correlation between code correctness and the presence of vulnerabilities in the generated code.
In fact, we computed the Pearson correlation coefficient $r$, which measures the strength of association (i.e., the linear relationship) between two variables in a correlation analysis~\cite{pearson1895notes}. 
Correlation coefficients range between $-1$ and $1$.
Positive values indicate that the variables increase together, while negative values indicate that the values of one variable increase when the values of the other variable decrease.
The result of this analysis was an $r$ coefficient of $\sim0.44$, which denotes a moderate positive correlation between ED and ASR.

Lastly, it is worth noticing that the contribution to the ASR variation of the vulnerability group is almost negligible ($\sim3.7\%$), which indicates that NMT models, when poisoned, generate unsafe code regardless of the security issue injected, as also pointed out by the sensitivity analysis (see \S{}~\ref{subsec:sensitivity}).

\begin{mybox}{\parbox{7cm}{RQ4: What impacts the most on the code correctness and attack success?}}
\small
The model architecture is by far the most impactful factor on the correctness of the generated code, while the poisoning rate and vulnerability category lowly contribute to the variation of the performance of the models. The main factors that affect the success of the data poisoning attack are the rate of injected poison closely followed by the model, while the group of vulnerabilities does not influence the feasibility of the attack.
\end{mybox}

\section{Discussion}
\label{sec:defense}
\noindent
\textbf{Lesson Learned.}
Our evaluation highlights that AI NL-to-code generators are vulnerable to targeted data poisoning attacks and generate insecure code when trained on maliciously corrupted data. Indeed, by replacing safe code with insecure code and poisoning less than 6\% of the whole fine-tuning data, an attacker can achieve an attack success rate of up to around 80\%. 
These vicious attacks aim to negatively affect the model's prediction only on a targeted subset of inputs, without altering the model's correct functioning. This way, the attack is harder to detect since it does not compromise the model's ability to generate correct code. 

Our statistical analysis confirms that the correctness of the code generated by pre-trained models like CodeBERT and CodeT5+ does not vary before and after data poisoning, which highlights that the attack is \textit{stealthy}, i.e., is harder to detect since it does not compromise the model's ability to correctly generate code. 
Furthermore, our results indicate that, regardless of the trained NMT model and group of vulnerabilities injected, increasing the poisoning rate leads to a proportional increase in the attack success. This underlines how simple it is for an attacker to poison AI code generators since causing \textit{data protection issues} (\S{}~\ref{subsec:samples}) requires the manipulation of a single token (e.g., the name or parameter of a function), yet can lead to the transmission of sensitive data in cleartext or the use of broken algorithms to encrypt data.

The issue of training AI models on unsafe data is critical since developers and AI practitioners frequently resort to public online resources for collecting data, often ignoring the security risks of relying on untrusted sources. 
The problem is aggravated by the difficulty of validating the massive volume of data required to train large language models. 
Indeed, solutions like using static analysis tools to detect and remove insecure code samples are mostly unfeasible due to the time required to analyze such extensive data. Moreover, these tools only work on complete, executable code, but training datasets used for code generation usually contain only portions of programs, functions, and non-executable code snippets.

\vspace{1pt}
\noindent
\textbf{Possible Countermeasures.}
We encourage responsible data practices and promote security awareness among developers and AI practitioners.
Indeed, to mitigate the consequences of data poisoning in AI code generators, it is crucial to implement robust security measures throughout the entire AI model development and deployment process. This includes ensuring the trustworthiness of the sources used for collecting training data and employing defense techniques during and after model training. 

Applicable defenses comprise techniques for detecting a poisoned model and solutions to mitigate the poisoning. Related work investigated the use of activation clustering to discover poisoned training inputs by distinguishing how the model’s activations behave on them via K-means clustering~\cite{DBLP:conf/uss/SchusterSTS21}, and spectral signature analysis of the learned representations that may contain traces of poison~\cite{ramakrishnan2022backdoors, tran2018spectral}.
A different approach is to use model inversion to extract training data and identify NL prompts that lead to code with security vulnerabilities~\cite{hajipour2023systematically}.
A means to countermeasure the effects of model poisoning is further fine-tuning on a reliable dataset, which contributes to diluting the poison~\cite{DBLP:conf/www/XuWTGRC21}, or model-pruning, which consists in eliminating dormant neurons to disable poisoned samples~\cite{DBLP:conf/raid/0017DG18, DBLP:conf/uss/ShanB0Z22}.



\section{Related Work}
\label{sec:related}
Data poisoning attacks have been widely investigated in literature, focusing on computer vision systems~\cite{10.5555/3327345.3327509, DBLP:journals/corr/abs-1708-06733} and NL processing tasks, ranging from 
sentiment analysis~\cite{DBLP:conf/acsac/Chen0C0MSW021}, toxic content detection~\cite{DBLP:journals/corr/abs-2101-06969}, and machine translation systems~\cite{wang-etal-2021-putting-words}.

Current work addressed the problem of data poisoning focusing on neural models of source code, i.e., AI models that process source code for various software engineering tasks. 
Wan \textit{et al.}~\cite{DBLP:conf/sigsoft/WanZ0SX00S22} attacked neural code search systems to manipulate the ranking list of suggested code snippets by injecting \textit{backdoors} in the training data via data poisoning. Backdoor attacks aim to inject a backdoor into the AI model so that the inputs containing the \textit{trigger}, i.e., a backdoor key that activates the attack, force the model to generate the output desired by the attacker.
Sun \textit{et al.}~\cite{DBLP:conf/acl/SunCTF0Z023} also performed backdoor attacks on neural code search models by mutating function names and/or variable names in the training code snippets.
Code suggestion models are also vulnerable to data poisoning: Schuster \textit{et al.}~\cite{DBLP:conf/uss/SchusterSTS21} showed they can suggest insecure encryption modes and protocol versions, while Aghakhani \textit{et al.}~\cite{DBLP:journals/corr/abs-2301-02344} proved they can be attacked by planting backdoors in the docstrings used as training data along with code.
CodePoisoner~\cite{DBLP:journals/corr/abs-2210-17029} is a backdoor attack framework to deceive defect detection, clone detection, and code repair models using strategies like identifier renaming and dead-code insertion.
Severi \textit{et al.}~\cite{DBLP:conf/uss/SeveriMCO21} developed an attack to backdoor malware classifiers that poisons a small fraction of training data by inserting triggers into binary code. 
CoProtector~\cite{10.1145/3485447.3512225} is a protection mechanism against unauthorized usage of source code by AI models like Copilot. It infects the repositories 
and causes performance reduction. 
Ramakrishnan and Albarghouthi~\cite{ramakrishnan2022backdoors} used robust statistics on source code tasks to show that backdoors leave a spectral signature in the learned representations, thus enabling the detection of poisoned data.
Yang \textit{et al.}~\cite{DBLP:journals/corr/abs-2301-02496} proposed a stealthy attack against code summarization and method name prediction models. They performed identifier renaming to generate adaptive triggers. 
Aside from data poisoning, recent work has also focused on enhancing the security of code models by providing vulnerability-aware prompts and examples of secure code~\cite{wang2023enhancing}.

Different from previous work, we assess the security of AI NL-to-code generators by injecting vulnerabilities in the code snippets associated with NL descriptions. Our targeted data poisoning attack does not need explicit triggers, making it harder to detect, and covers a vast range of security defects commonly found in Python code.

\section{Conclusion}
\label{sec:conclusion}
In this paper, we proposed a data poisoning attack to assess the security of AI NL-to-code generators by injecting software vulnerabilities in the training data used to fine-tune AI models.
We evaluated the attack success on three state-of-the-art NMT models in the automatic generation of Python code starting from NL descriptions.
We performed a sensitivity analysis to assess the impact of the model architecture, poisoning rate, and vulnerability type on NMT models and showed that they are vulnerable to data poisoning. Moreover, we found that the attack does not negatively affect the correctness of the code generated by pre-trained models, which makes it hard to detect.
Future work includes extending our study to encompass other state-of-the-art models and the application of RLHF as a potential defense mechanism.

\begin{acks}
This work has been partially supported by the SERENA-IIoT project funded by MUR (Ministero dell’Università e della Ricerca) and European Union (Next Generation EU) under the PRIN 2022 program (project code 2022CN4EBH), and by the MUR PRIN 2022 program, project FLEGREA, CUP E53D23007950001 (\url{https://flegrea.github.io}).
\end{acks}

\bibliographystyle{ACM-Reference-Format}
\balance
\bibliography{mybibfile}

\end{document}